# Eddy Covariance: A Scientometric Review (1981-2018)

**Authors**: Tian-Yuan Huang[1], Yi-Fei Liu[1], Yuan-Chen Wang[1], Hai-Qing Guo[1], Jun Ma[1], Bin Zhao[1*]

**Affiliations**:

1 Ministry of Education Key Laboratory for Biodiversity Science and Ecological Engineering, Coastal Ecosystems Research Station of the Yangtze River Estuary, and Shanghai Institute of EcoChongming (SIEC), Fudan University, Shanghai 200433, China

**Contact Information**:
*Bin Zhao (corresponding author)
Email: zhaobin@fudan.edu.cn

**Key points:**

1. A comprehensive, systematic and critical scientometric analysis on eddy covariance measuring system was carried out.

2. Research trends in the ~40 years has been showed and discussed.

3. Performance of FLUXNET has been quantified and evaluated.

# Abstract

The history of eddy covariance (EC) measuring system could be dated back to 100 years ago, but it was not until the recent decades that EC gains popularity and being widely used in global change ecological studies, with explosion of related work published in papers from various journals. Investigating 8297 literature related with EC from 1981 to 2018, we make a comprehensive and critical review of scientific development of EC from a scientometric perspective. First, the paper outlines general bibliometric statistics, including publication number, country contribution, productive institutions, active authors, journal distribution, highly cited articles and fund support, to provide an informative picture of EC studies. Second, research trends are revealed by network visualization and modeling based on keyword analysis, from where we could discover the knowledge structure of EC and detect the research focus and hotspots transitions at different periods. Third, collaboration in EC research community has been explored. FLUXNET is the largest global network uniting EC researchers, here we have quantified and evaluated its performance by using bibliometric indicators of cooperation and citation. Specific discussions have been given to the historical development of EC, including technical maturation and application promotion. Considering the current barrier for collaboration, the review closes by analyzing the reasons hindering data sharing and makes a prospect of new models for data-intensive collaboration in the future.

# 1. Introduction

Providing the most direct measures of trace gas and energy exchange between Earth's surface and atmosphere(Baldocchi 2014), eddy covariance is a powerful technique for ecologists to investigate the energy flow and footprint of gaseous compounds ($CO_2$, $CH_4$, $H_2O$, $N_2O$, etc.). Flux values calculated by statistical methods is a time series data revealing exchange at ecosystem scale covering large areas(Hill et al. 2017). With its great strength and rapid development, eddy covariance has become a routine standard method to assess ecosystem carbon exchange and beyond.

Recent four decades have seen a boom in scientific studies based on eddy covariance, massive papers have been published recording the progress of technology, transition of hotspots and advancement of cognition. It would be impossible and unnecessary for researchers to read all of them. While conventional literature reviews still serve as indispensable sources for scholars to learn a research field quickly and keep themselves familiar with the frontiers, in the information age scientometrics, quantitative study of science, technology and innovation(Mingers and Leydesdorff 2015), is offering scientists an alternative new approach to get the overview of their interested study area. Armed with well documented literature database and sound data science workflow, scientometrics has now been widely used for research evaluation in various science

fields(Jappe 2007, Liu et al. 2011, Wang et al. 2018, Young and Wolf 2006, Zhuang et al. 2013).

With the recent rapid progress of eddy covariance studies and its increasingly significant role in biogeoscience and global change science, it is important for researchers to reflect on the development track and manage their future study plans frequently. Research reviews had been made for eddy covariance studies(Baldocchi 2014, Baldocchi 2003, Reichstein et al. 2005, Velasco and Roth 2010), however, few of them were designed from the perspective of scientometrics, which provides deep and nuance understanding via systematic mapping and bibliometrics in the framework of research weaving (Nakagawa et al. 2018). Therefore, here we carry out a comprehensive and systematic review considering the historical development, current state and future prospects of eddy covariance studies. In summary, our work has applied various quantitative scientometric analysis on bibliometric data of eddy covariance to offer: (1) General view of eddy covariance studies based on bibliometric information, including publication number, countries, institutions, authors, journals, citations and funds. (2) Change of research hotspots in eddy covariance studies, applying both visualization and modelling in network science supplemented by citation analysis. (3) Collaborative status in eddy covariance research community and performance of collaborative project FLUXNET based on citation and cooperation. Hopefully, our scientometric review could provide researchers with new and critical insights in knowledge advancement, technology development, collaboration establishment and scientific policy-making in various study fields supported by eddy covariance technique.

## 2. Materials and methods

### 2.1 Data source

The main data of bibliography were collected from Scopus database (https://www.scopus.com). We have referred to our previous study(Dai et al. 2018) to determine the synonymous keywords of "eddy covariance", and a total of 8297 publications ranged from 1981 to 2018 were searched under advanced mode using the following formula:

TITLE-ABS-KEY ({eddy flux} OR {eddy-flux} OR {eddy fluxes} OR

{eddy covariance} OR {eddy-covariance} OR

{eddy correlation} OR {eddy-correlation} OR

{flux tower} OR {flux towers}) AND

PUBYEAR > 1980 AND PUBYEAR < 2019

The retrieval date was June 2, 2019 (this would affect the total number of citation count). R package "rscopus" was used to retrieve the data(Muschelli 2018). These records include the information of title, abstract, keywords, references, author, institution, etc. In addition, the bibliography information of publications that cited these 8297 papers were also retrieved in Scopus database. The annual citation count of papers after publication was retrieved in Web of Science (http://apps.webofknowledge.com). Data management and presentation were mainly carried out using 'tidyverse'(Wickham 2016) package under R environment(R Core Team 2013). The geographic distribution map was drawn using 'sf'(Pebesma 2018) and 'rnaturalearth'(South 2017). Visualization of tree map was realized in 'treemapify'(Wilkins 2017), while the temporal trends laid in tables were drawn based on R packages of 'DT'(Xie et al. 2015) and 'sparkline'(Vaidyanathan et al. 2016).

## 2.2 H-index and CiteScore

In our study, we used H-index to evaluate the influence of authors and institutions, while CiteScore was applied to estimate the impact of journals. Proposed by Hirsch in 2005, H-index was first defined as "the number of papers with citation number $\geq$

h"(Hirsch 2005). In detail, H-index is "the highest number of publications of a scientist that received h or more citations each while the other publications have not more than h citations each"(Schreiber 2008). Based on citation counts of publications, it has been widely used to evaluate the scholars' scientific research output, and could be extended to measure the impact of journals, institutions and countries as well. In our study, we have used H-index to evaluate the influence of authors and institutions.

CiteScore is an index created by Elsevier on December 8, 2016 (https://www.elsevier.com/editors-update/story/journal-metrics/citescore-a-new-metric-to-help-you-choose-the-right-journal). Similar to Impact Factor (IF), CiteScore tracked performance of journals according to the number of citations over a period of time. However, CiteScore uses a time window of 3 years instead of 2 in IF. Moreover, the databases that support CiteScore and IF are different. Main features of CiteScore and its comparison with IF could be found in previous work(Da Silva and Memon 2017). Because the major database of our research is from Scopus, here we used 2017 CiteScore™ values (https://www.elsevier.com/about/press-releases/science-and-technology/elsevier-releases-2017-citescore-values) to evaluate the impact of journals. The calculation method could be found as below:

$$\text{CiteScore}^{\text{TM}}_{2017} = \frac{\text{number of citations in 2017 to articles published in 2016,2015, and 2014}}{\text{number of articles published in 2016,2015, and 2014}}$$

## 2.3 Network analysis and visualization

Network analysis and visualization has mainly been applied in two sections in our investigation. One is in the exploration of historical trends of eddy covariance studies, where we have implemented a keyword analysis based on their co-occurrence relations. The two maps of keywords visualized as networks were created in VOSviewer(Waltman 2009), displaying the clustering results and temporal transition of topics respectively. In the map showing clustering results, the label size and circle size were determined by the weights of keywords, which is the total count of keyword occurrence. Only keywords occurring more than 20 times during the investigated period were showed. Keywords from different clusters were displayed in different colors, and the links between keywords were showing the co-occurrence relations among them. The thicker the links, the more likely that the connecting two keywords would co-occur with each other. Applying the overlay visualization, the same information could be used to reflect the temporal trends of topics according to the occurring year of these keywords. The brighter the color, the more likely that these keywords were occurred in recent years. Details of working principles, functions and operation methods could be found on the official website at http://www.vosviewer.com/.

Another implementation of network analysis and visualization was applied to explore the effect of FLUXNET on collaboration among eddy covariance researchers at three levels, namely individual level, institutional level and international level. First, we must recognize the FLUXNET research from the overall research supported by eddy covariance. We assumed that if the researchers have used resources from the research networks, they would mention the names of networks in the fields of title, abstract or keywords. Therefore, we added restricted conditions to the previous formula to find these entries according to the names of research networks in FLUXNET according to https://fluxnet.fluxdata.org/about/regional-networks/. The final constructed formula is showed as below:

TITLE-ABS-KEY ({eddy flux} OR {eddy-flux} OR {eddy fluxes} OR {eddy covariance} OR {eddy-covariance} OR {eddy correlation} OR {eddy-correlation} OR {flux tower} OR {flux towers}) AND

TITLE-ABS-KEY ({AmeriFlux} OR {AsiaFlux} OR {Boreal Ecosystem Research and Monitoring Sites} OR {Canadian Carbon Program} OR {CarboAfrica} OR {CarboEurope} OR {CarboItaly} OR {Carbomont} OR {ChinaFlux} OR {China Flux}

OR {EuroFlux} OR {European Fluxes Database} OR {Fluxnet-Canada} OR {GreenGrass} OR {Integrated Carbon Observation System} OR {Infrastructure for Measurements of the European Carbon Cycle} OR {Inland Water Greenhouse Gas FLUX} OR {JapanFlux} OR {KoFlux} OR {The Large Scale Biosphere-Atmosphere Experiment in Amazonia} OR {MexFlux} OR {Nordic Centre for Studies of Ecosystem Carbon Exchange} OR {OzFlux} OR {RusFluxNet} OR {Swiss Fluxnet} OR {Terrestrial Carbon Observation System Siberia} OR {Urban Fluxnet} OR {US-China Carbon Consortium} OR {Fluxnet}) AND PUBYEAR > 1980 AND PUBYEAR < 2019

Using the above formula, 680 entries of publications were retrieved at June 2, 2019. These publications are a subset of our overall database, and could be regarded as studies supported by research networks of eddy covariance, we would call them FLUXNET research in our investigation.

Social network analysis is frequently used to depict the collaboration among scientists, institutions and countries. In our case, the nodes in networks represent authors, institutions or countries. Take individual level as an example (which could be extended to institutional and international level), if a paper has more than one authors, they would be connected to each other by links to form a small network. Summarizing numerous papers could merge these small networks to a large one. By analyzing and visualizing the structure of the network with or without FLUXNET research, we could learn better about the effect of FLUXNET on the collaboration behavior in the research community. For networks excluding or including FLUXNET research, R package 'netbiov'(Tripathi et al. 2014) was applied for visualization. In this way, we could show all the nodes in the network for comparison. Nodes in the same module would be merged into one big node based on fastgreedy algorithm. In addition, indicator from network science, averaged degree (AD) was calculated to evaluate the structure of networks. In network science, degree of a node is defined as the links it has with other nodes. Average degree is the mean degree of all nodes in a network, showing connectivity of nodes. For instance, the AD of author collaboration network could tell us how well these individuals were connected with each other. The network analysis for this section was carried out using R packages of 'igraph'(Csardi and Nepusz 2006) and 'tidygraph'(Pedersen 2018).

## 2.4 Research hotspots detection

Keyword frequency is commonly used to detect research hotspots in a field. Our

previous study, however, suggested that simply using keyword frequency would cause bias because it does not take keyword relations, research cycle and temporal trends into consideration(Huang and Zhao 2019). Therefore, we utilized a new method named as 'PAFit'(Pham et al. 2016) to measure the popularity of keywords in a temporal dynamical network. In this method, the keyword co-occurrence network is growing over time, and follows the principles of "rich get richer" and "fit get richer". This method has been proposed and verified in our previous research, and is applied to calculate the popularity index (PI) and fitness index (FI) in our current study. PI could reflect the popularity of a keyword, it has the power to estimate the probability of a keyword to occur in a paper or co-occur with other keywords in the future. On the other hand, FI could depict the temporal trend of the keywords, hence is used to detect potential hotspots in our study. R package 'PAFit'(Pham et al. 2017) was applied for the calculation of these two indexes.

## 3. General view of eddy covariance studies

### 3.1 Overall trends of publication outputs

Generally, the publication outputs supported by eddy covariance is on the rising trend (Fig. 1). Before 21$^{st}$ century, the speed of development was slow. In 1990, only 32 publications are tracked in Scopus database contributed by 71 authors in total. While in 2000, 122 publications could be tracked contributed by 389 authors. A burst could be found in the first decade of 21[st] century. Within the period of 2001-2010, the number of publications grew by 29.9 per year. By 2010, there were 408 publications contributed by 1609 authors. 1459 institutions from 78 countries have carried out study based on eddy covariance technique in 2010, while these numbers were 494 and 42 respectively back in 2000. This trend maintained a good momentum after 2010. Up to 2018, we could find 8297 publications in total contributed by 15,860 distinct authors. Altogether, 3152 institutions from 103 countries have conducted their research based on eddy covariance from 1981 to 2018. Moreover, as time went by, studies using eddy covariance technique were reported by more journals covering a broad range of topics, by 2018 we could identify 927 journals containing works using this technique. For publications with funds recognizable, 81.8% papers were published in 2011-2018, while only 18.2% were published in the prior three decades. In early days, eddy covariance technique was not mature enough to put into use. With theoretical framework established before 1900(Reynolds 1895), explorations were unavoidable and mistakes were waiting to be made before more accurate measurements could be carried out(Dyer and Pruitt 1962, Dyer et al. 1967, Raupach 1979, Swinbank 1967).

Nevertheless, as performance of sensor improved, measurements of flux could be conducted for longer period at shorter intervals(Massman and Lee 2002, Wofsy et al. 1993). In addition, the big data era has facilitated the acquisition and management of flux data and induce scientists to collaborate and form networks of sites(Baldocchi 2003), which lifted the power of eddy covariance greatly.

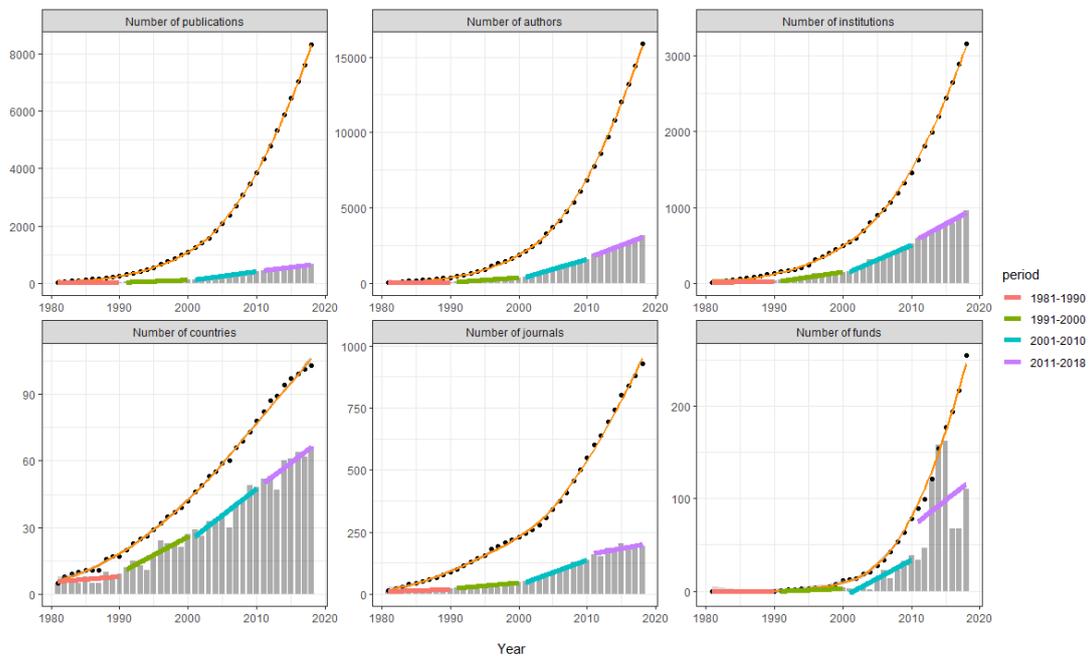

**Fig. 1. Overall trends of publication outputs from 1981 to 2018. The orange line represents the cumulative sum by time. The bars represent the annual number with colorful lines (namely red, green, blue and purple) showing the periodical trends using simple linear model.**

## 3.2 Contribution by countries

Number of publications for a given country could reflect its research strength and recognition degree in a specific field. According to statistical summary, there were 103 countries participate in research supported by eddy covariance technique from 1981 to 2018. Contribution of countries was estimated by the institution locations of the authors, and it should be noted that in our study "United Kingdom" refers to England, Scotland, Wales and North Ireland; "United States" refers to United States of America; "China" refers to mainland China, documents of Hong Kong and Taiwan are not included under "China" but analyzed as separate entities.

From the geographical distribution of publications (Fig. 2), it could be found that North America, Western Europe, Eastern Asia and Oceania were main regions that carry out most research based on eddy covariance. Among all the countries, United States leads the productivity rankings with 3302 publications in total, taking 39.8% of the whole share (Table 1). China comes second with 1525 publications, followed by Germany, Canada, United Kingdom, France, Japan, Italy, Netherlands and Australia in a descending order. Focus on the temporal trends (Fig. 3), we could find that US took the head early at the first decade since 1980, and keep the place all the way through the 38 years. On the other hand, only after 2000 did China have more publication based on eddy covariance. Nevertheless, China maintained the momentum of growth and leapt to the second place within the first decade of 21$^{st}$ century. At the same period, the publications of other countries were all increasing. More countries are contributing as eddy covariance is getting more popular as a technique to measure fluxes and applied to various circumstances, it could be expected that the distribution of countries on publication would be more balanced in the future.

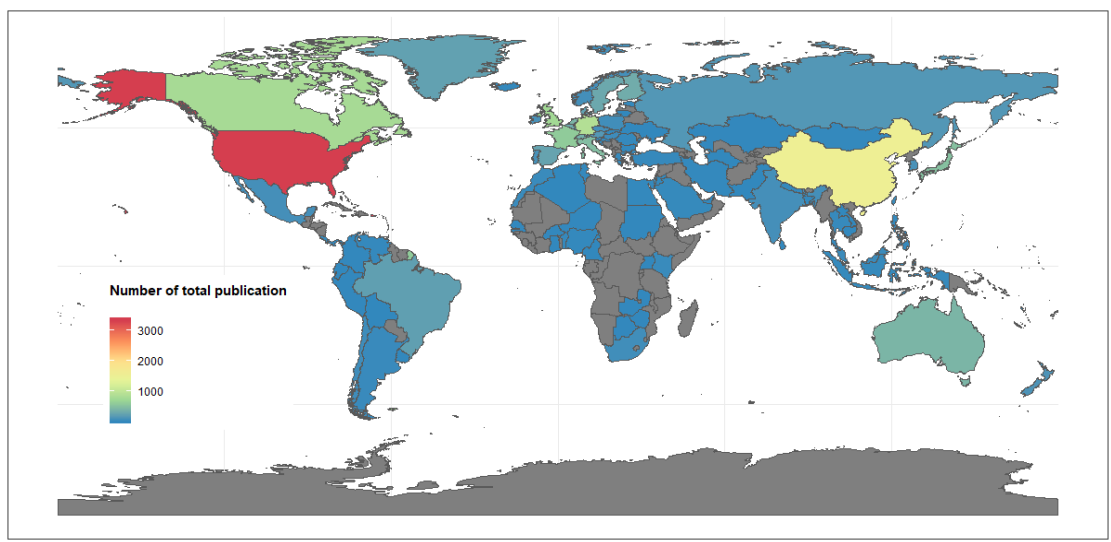

**Fig. 2. Geographical distributions of publications, 1981-2018.**

**Table 1 Top 10 productive countries, 1981-2018**

| Rank | Country | TP | TP R(%) |
|---|---|---|---|
| 1 | United States | 3302 | 39.8 |
| 2 | China | 1525 | 18.38 |
| 3 | Germany | 874 | 10.53 |
| 4 | Canada | 774 | 9.33 |
| 5 | United Kingdom | 731 | 8.81 |
| 6 | France | 579 | 6.98 |
| 7 | Japan | 487 | 5.87 |
| 8 | Italy | 463 | 5.58 |
| 9 | Netherlands | 398 | 4.8 |
| 10 | Australia | 397 | 4.78 |

Note: TP is the number of total publications; TP R (%) is the ratio of the number of documents contributed by a country to the total number of publications in 1981-2018.

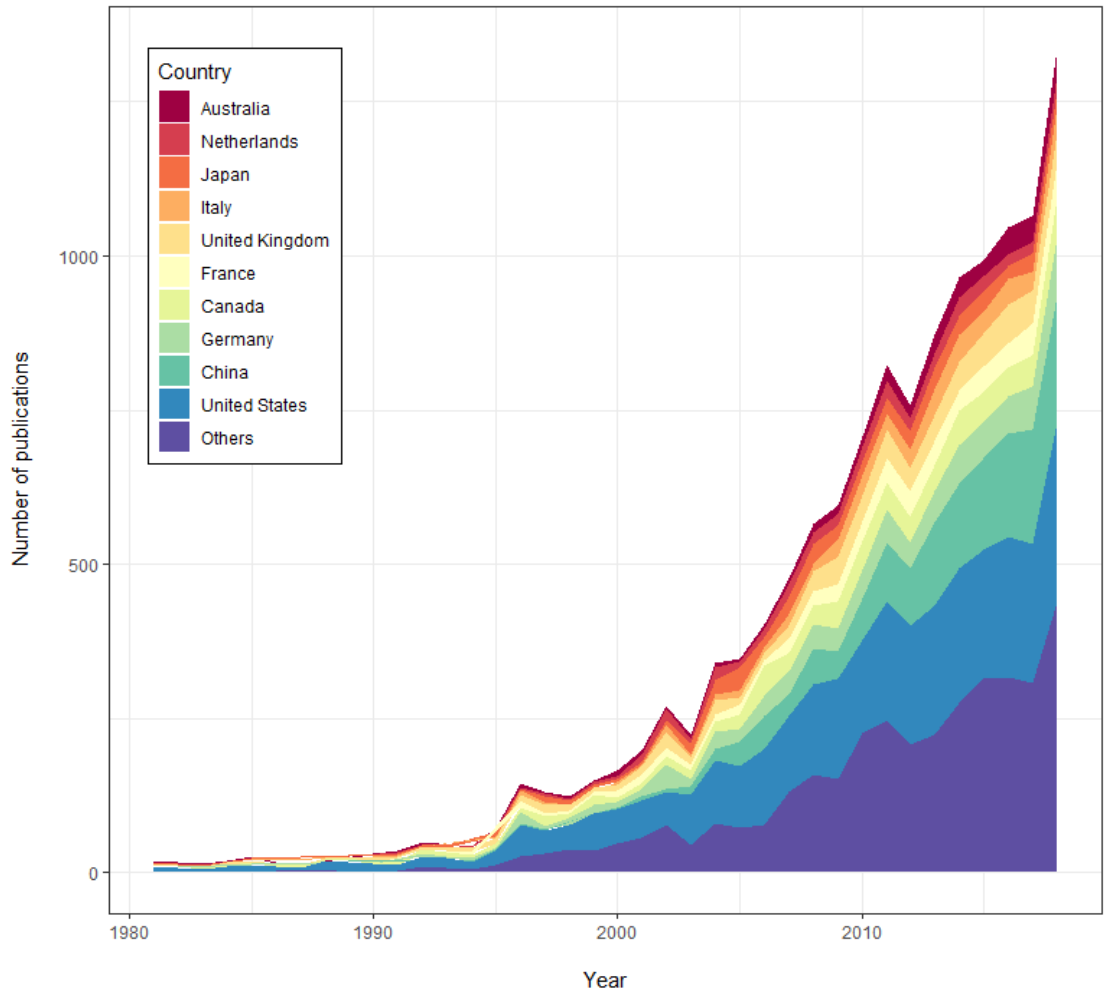

**Fig. 3. Trends of publications in most productive countries, 1981-2018.**

## 3.3 Productive institutions

Table 2 listed the top 15 research institutions with most publications supported by eddy covariance. Among these institutions, six of them come from United States, Canada and United Kingdom each has two, while China, Finland, Germany, Netherlands, Sweden and Switzerland each hold one institution respectively. These institutions are all from developed countries, except for China which emerges as booming economy in recent years. The distribution of institutes is wide across the globe, indicating that eddy covariance has been widely accepted as a practical technique in the world. In term of total publications, Institute of Geographical Sciences and Natural Resources Research, China Academy of Sciences was tracked as the institute with most publications supported by eddy covariance from 1981 to 2018, with 311 papers contributing to 20.39% of the total number of Chinese documents. Considering H-index, Harvard University has taken the first place. It has 75 publications with at least 75 citations before 2019. Commonalities could be found from high-frequency keywords of these institutions. Evapotranspiration (ET) and net ecosystem exchange (NEE) are hotspots of research supported by eddy covariance. Meanwhile, some institutes would have their own research focus and features. For instance, Wageningen University and Research Centre has rich and extensive experience on using scintillometer to quantify turbulent characteristics of atmosphere(De Bruin et al. 1995, Jia et al. 2003, Meijninger et al. 2002), while Lund University does well in climate modeling at peatlands(Gallego-Sala et al. 2018).

**Table 2 Top 15 productive institutes (ranked by total publications), 1981-2018**

| Rank | Institution | Country | TP | TT | TP RC(%) | TP RW(%) | H-index | HFK |
|---|---|---|---|---|---|---|---|---|
| 1 | Institute of Geographical Sciences and Natural Resources Research, Chinese Academy of Sciences | China | 311 | | 20.39 | 3.75 | 37 | evapotranspiration, modis, gross primary production |
| 2 | Helsingin Yliopisto | Finland | 247 | | 77.43 | 2.98 | 57 | net ecosystem exchange, photosynthesis, ecosystem respiration |
| 3 | University of Colorado at Boulder | United States | 229 | | 6.94 | 2.76 | 63 | net ecosystem exchange, evapotranspiration, ecosystem respiration |
| 4 | Max Planck Institut für Biogeochemie Jena | Germany | 216 | | 24.71 | 2.6 | 68 | net ecosystem exchange, evapotranspiration, carbon balance |
| 5 | The University of British Columbia | Canada | 213 | | 27.52 | 2.57 | 62 | evapotranspiration, photosynthesis, net ecosystem productivity |
| 6 | Oregon State University | United States | 199 | | 6.03 | 2.4 | 66 | ecosystem respiration, net ecosystem exchange, evapotranspiration |
| 7 | ETH Zürich | Switzerland | 196 | | 57.48 | 2.36 | 54 | drought, ecosystem respiration, gross primary production |
| 7 | Harvard University | United States | 196 | | 5.94 | 2.36 | 75 | modis, phenology, gross primary production, net ecosystem exchange, photosynthesis |
| 9 | University of California, Berkeley | United States | 186 | | 5.63 | 2.24 | 72 | evapotranspiration, net ecosystem exchange, evaporation |
| 10 | Wageningen University and Research Centre | Netherlands | 180 | | 45.23 | 2.17 | 60 | scintillometer, gross primary production, net ecosystem exchange |
| 11 | Centre for Ecology & Hydrology | United Kingdom | 176 | | 24.08 | 2.12 | 54 | evaporation, ecosystem respiration, grassland |
| 12 | National Oceanic and Atmospheric Administration | United States | 165 | | 5 | 1.99 | 57 | carbon dioxide, carbon balance, evaporation, evapotranspiration |
| 13 | University of Edinburgh | United Kingdom | 164 | | 22.44 | 1.98 | 60 | carbon balance, carbon dioxide, data assimilation, energy balance, evapotranspiration |
| 14 | Lunds Universitet | Sweden | 163 | | 57.39 | 1.96 | 44 | net ecosystem exchange, peatland, remote sensing |
| 15 | Environment Canada | Canada | 161 | | 20.8 | 1.94 | 53 | boreal forest, photosynthesis, evapotranspiration |
| 15 | USDA Forest Service | United States | 161 | | 4.88 | 1.94 | 64 | evapotranspiration, ecosystem respiration, net ecosystem exchange |

Note: TP is the number of total publications; TT is the temporal trends; TP RC[W] (%) is the ratio of publications to the whole country[world]; HFK is high-frequency keywords excluding the searched keywords ("eddy covariance", etc.).

## 3.4 Active authors

As a growing technology, eddy covariance has attracted more and more researchers to do investigations on this method and apply it in their studies. From 1981 to 2000, there are 1890 distinct authors tracked as doing their research based on eddy covariance. This number has raised to 5619 at the period of 2001-2010. Table 3 shows a list of active authors who published most papers using eddy covariance. Among these researchers, seven of them come from United States, while Canada and China have two respectively, and Finland, France, Germany and Sweden each has one. The most productive authors were Black T. from University of British Columbia, Canada and Yu G. from Institute of Geographical Sciences and Natural Resources Research, China Academy of Sciences, each published 147 papers of eddy covariance tracked in Scopus database. Holding interests in biometeorology, soil physics and microclimate modification, Black T. is a pioneer scholar using eddy covariance to measure forest evapotranspiration, his early work could be traced back to 1981 in our investigation(Spittlehouse and Black 1981). On the other hand, Yu G. carried out extensive studies on ecosystem respiration, evapotranspiration, carbon dioxide ($CO_2$) flux, gross primary productivity and net ecosystem exchange(Yu et al. 2014, Yu et al. 2013). He is also the primary initiator of Chinese Terrestrial Ecosystem Flux Research Network (ChinaFlux), which fills important regional gap in FLUXNET(Yu et al. 2006). The author with highest total number of citations was Baldoccchi D. from Department of Environmental Science, Policy, & Management, UC Berkeley. For the tracked 132 papers in the field of eddy covariance, his works have been cited 22,307 times in total by 2018. From the temporal trends of publications, we could find that he has carried out his studies and published his work as early as 1985. He is also the author with highest H-index in the field, at least 73 of his publications have been cited at least 73 times during the investigated period. His main research focus is to "understand how terrestrial biosphere breathes", including the mechanisms of gas exchange between vegetation, soil and atmosphere(Baldocchi 2003, Baldocchi et al. 1988).

**Table 3 Top 15 active authors (ranked by total publications), 1981-2018**

| Rank | Author | Country | TP | TT | TC | CPP | H-index | HFK |
|------|--------|---------|-----|-----|-----|------|---------|-----|
| 1 | Black T. | Canada | 147 | 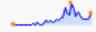 | 9492 | 64.57 | 57 | evapotranspiration, net ecosystem productivity, boreal forest, photosynthesis |
| 1 | Yu G. | China | 147 | 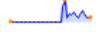 | 3879 | 26.39 | 32 | ecosystem respiration, evapotranspiration, co2 flux, gross primary productivity, net ecosystem exchange |
| 3 | Baldocchi D. | United States | 132 | 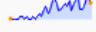 | 22307 | 168.99 | 73 | evapotranspiration, micrometeorology, carbon dioxide, evaporation |
| 3 | Vesala T. | Finland | 132 | 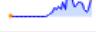 | 11060 | 83.79 | 49 | net ecosystem exchange, photosynthesis, evapotranspiration, gross primary production |
| 5 | Chen J. | United States | 107 | 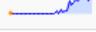 | 4910 | 45.89 | 39 | evapotranspiration, ecosystem respiration, gross primary production, modis, net ecosystem exchange |
| 6 | Wofsy S. | United States | 91 | 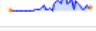 | 14263 | 156.74 | 57 | net ecosystem exchange, remote sensing, gross primary production, modis, respiration |
| 7 | Hollinger D. | United States | 86 | 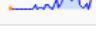 | 12958 | 150.67 | 53 | net ecosystem exchange, phenology, ecosystem respiration, modis, uncertainty |
| 8 | Sun X. | China | 81 | 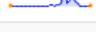 | 2257 | 27.86 | 27 | co2 flux, eddy covariance method, ecosystem respiration |
| 9 | Barr A. | Canada | 80 | 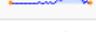 | 6474 | 80.92 | 44 | boreal forest, photosynthesis, carbon balance, drought, evapotranspiration |
| 9 | Katul G. | United States | 80 | 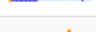 | 7584 | 94.8 | 45 | net ecosystem exchange, evapotranspiration, ecosystem respiration, pinus taeda |
| 9 | Lindroth A. | Sweden | 80 | 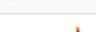 | 6014 | 75.17 | 38 | photosynthesis, respiration, carbon balance, net ecosystem exchange |
| 12 | Law B. | United States | 79 | 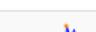 | 11096 | 140.46 | 45 | ecosystem respiration, net ecosystem exchange, modis, pinus ponderosa |
| 13 | Bernhofer C. | Germany | 77 | 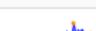 | 10800 | 140.26 | 39 | gross primary production, brook90, carbon budget, ecosystem respiration, forest ecosystems |
| 13 | Ciais P. | France | 77 | 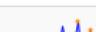 | 4806 | 62.42 | 33 | carbon cycle, climate change, carbon fluxes, grassland, land surface model |
| 13 | Kustas W. | United States | 77 | 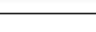 | 4218 | 54.78 | 32 | evapotranspiration, remote sensing, advection, energy balance |

Note: TP is the number of total publications; TT is the temporal trends; TC is the total citation count; CPP is citations per publication; HFK is high-frequency keywords.

## 3.5 Journal distribution

Journal distribution of research supported by eddy covariance is rather concentrated. Top 15 most productive journals have contributed to 45.2% of total publications from 774 journals (Table 4). According to Scopus source list (https://www.scopus.com/sources, retrieved at March, 2019), the main topics of these journals could be divided into three major categories: 1. Earth and Planetary Sciences; 2. Agricultural and Biological Science; 3. Environmental Science (Fig. 4). Most documents were published in journals covering the topic of Atmospheric Science (39.9%). The second largest source of publications based on eddy covariance was Forestry (22.1%), followed by Ecology (19.1%), Water Science and Technology (17.9%), Agronomy and Crop Science (16.5%) and Global Planetary Change (16.4%). In terms of most productive journals (Table 4), Agriculture and Forest Meteorology, Journal of Geophysical Research, Boundary-Layer Meteorology, Global Change Biology and Biogeosciences make the top 5 journals publishing most studies supported by eddy covariance. Reporting 935 studies (11.29% of total share) at our investigated period, Agriculture and Forest Meteorology has gained 49,025 citations from other documents, which takes the head in the subfield. Though holding a relatively less publications supported by eddy covariance, Global Change Biology has the highest citation per publication (101.82) among the top 15 journals in the list. Also, it has a high CiteScore of 9.12, indicating that in average its publication published from 2014 to 2016 has been cited 9.12 times by publications in 2017. It is worth noted that Journal of Geophysical Research, together with Boundary-Layer Meteorology and Journal of the Atmospheric Sciences, are journals that paid close attentions to eddy covariance technique at very early stages before 1990, publishing pioneering work of explorations on principles and methods. Only after 2000 did eddy covariance gain its popularity and attract more researches to use it for applications at various subfields such as agriculture, forestry, hydrology and ecology.

**Table 4 Top 15 productive journals, 1981-2018**

| Rank | Journal | TP | TT | TP R(%) | TC | CPP | CS |
|------|---------|-----|-----|---------|------|--------|------|
| 1 | Agricultural and Forest Meteorology | 935 | 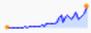 | 11.29 | 49025 | 52.43 | 4.67 |
| 2 | Journal of Geophysical Research | 688 | 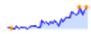 | 8.31 | 24600 | 35.76 | 3.19 |
| 3 | Boundary-Layer Meteorology | 305 | 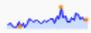 | 3.68 | 13066 | 42.84 | 2.47 |
| 4 | Global Change Biology | 282 | 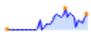 | 3.4 | 28714 | 101.82 | 9.12 |
| 5 | Biogeosciences | 260 | 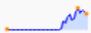 | 3.14 | 7171 | 27.58 | 3.96 |
| 6 | Remote Sensing of Environment | 158 | 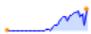 | 1.91 | 13107 | 82.96 | 7.16 |
| 7 | Journal of Hydrology | 148 | 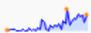 | 1.79 | 5775 | 39.02 | 4.06 |
| 8 | Atmospheric Environment | 142 | 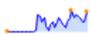 | 1.71 | 4352 | 30.65 | 4.12 |
| 9 | Geophysical Research Letters | 141 | 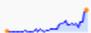 | 1.7 | 4381 | 31.07 | 4.51 |
| 10 | Atmospheric Chemistry and Physics | 135 | 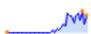 | 1.63 | 3688 | 27.32 | 5.44 |
| 11 | Journal of the Atmospheric Sciences | 135 | 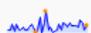 | 1.63 | 3954 | 29.29 | 3.13 |
| 12 | Journal of Physical Oceanography | 122 | 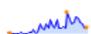 | 1.47 | 5251 | 43.04 | 2.99 |
| 13 | Hydrological Processes | 102 | 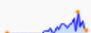 | 1.23 | 2766 | 27.12 | 3.15 |
| 14 | Agricultural Water Management | 97 | 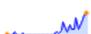 | 1.17 | 2437 | 25.12 | 3.49 |
| 15 | Remote Sensing | 95 | 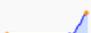 | 1.15 | 850 | 8.95 | 4.03 |

Note: TP is the number of total publications; TT is the temporal trends; TP R (%) is the ratio of the number of documents contributed by a journal to the total number of publications in 1981-2018; TC is the total citation count; CPP is citations per publication; CS is the CiteScore (2017) of a journal.

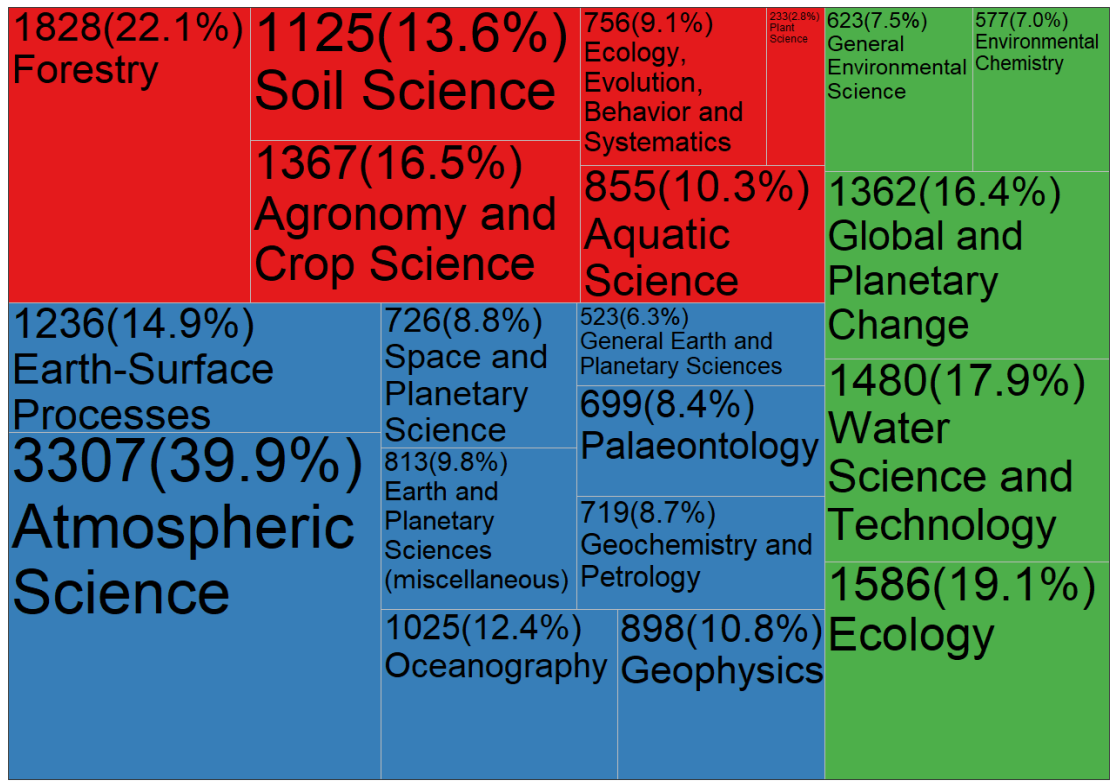

**Fig. 4. Top 20 productive subjects, 1981-2018. The number outside the brackets is number of total publications classified as the sub-subject, while the percentage number inside is the proportion of publications classified as the sub-subject to the total publication.**

### 3.6 Highly cited articles

Of the 8297 publications retrieved in our investigation, 566 were cited more than 100 times by June, 2019. For top 20 most-cited articles supported by eddy covariance (Table 5), Agricultural and Forest Meteorology has published five of them(Falge et al. 2001, Foken and Wichura 1996, Law et al. 2002, Twine et al. 2000, Wilson et al. 2002), whereas Global Change Biology has included four(Baldocchi 2003, Goulden et al. 1996,

Janssens et al. 2001, Reichstein et al. 2005). Remote Sensing of Environment and Science each has three(Mu et al. 2011, Mu et al. 2007, Zhao et al. 2005), while Boundary-Layer Meteorology contains two(McMillen 1988, Moore 1986). For the remaining, Annual Review of Fluid Mechanics(Finnigan 2000), Australian Journal of Botany(Baldocchi 2008) and Biogeosciences(Papale et al. 2006) each adopts one.

The earliest study in the list was published in 1986 on Boundary-Layer Meteorology titled as "Frequency response corrections for eddy correlation systems"(Moore 1986). This article displayed the effects of sensor response, path-length averaging, sensor separation and signal processing on eddy covariance systems frequency response, which provides invaluable methodological basis for the future study. Also focus on methodology, the article with most citations was published on Global Change Biology in 2005, titled as "On the separation of net ecosystem exchange into assimilation and ecosystem respiration: review and improved algorithm"(Reichstein et al. 2010). It has gained 1457 citations with an average annual citation count of 112.08. This research has reviewed on various methods that decompose net ecosystem exchange (NEE) into its main components, and suggested a new algorithm to avoid bias caused by confounding factors in seasonal data. Considering citation per year, article titled as "Improvements to a MODIS global terrestrial evapotranspiration algorithm" published on Remote Sensing of Environment in 2011 takes the lead with an average annual citation count of 136.43(Mu et al. 2011). This research team first developed a global remote sensing ET algorithm in 2007, which is also a highly cited work in eddy covariance studies(Mu et al. 2007). Then they improved this algorithm considering various factors, including vegetation cover faction, soil heat flux, stomatal conductance, etc. With higher accuracy, this work makes a huge contribution to generate global ET products ever after.

**Table 5 Top 20 most frequently cited articles, 1981-2018**

| Rank | Authors | Year | TC | CPY | Journal |
|------|---------|------|------|--------|---------|
| 1 | Reichstein et al. (2005) | 2005 | 1457 | 112.08 | Global Change Biology |
| 2 | Wilson et al. (2002) | 2002 | 1374 | 85.88 | Agricultural and Forest Meteorology |
| 3 | Baldocchi (2003) | 2003 | 1315 | 87.67 | Global Change Biology |
| 4 | Falge et al. (2001) | 2001 | 1138 | 66.94 | Agricultural and Forest Meteorology |
| 5 | Twine et al. (2000) | 2000 | 1060 | 58.89 | Agricultural and Forest Meteorology |
| 6 | Goulden et al. (1996) | 1996 | 1010 | 45.91 | Global Change Biology |
| 7 | Beer et al. (2010) | 2010 | 1009 | 126.12 | Science |
| 8 | Foken et al. (1996) | 1996 | 965 | 43.86 | Agricultural and Forest Meteorology |
| 9 | Mu et al. (2011) | 2011 | 955 | 136.43 | Remote Sensing of Environment |
| 10 | Finnigan (2000) | 2000 | 892 | 49.56 | Annual Review of Fluid Mechanics |
| 11 | Moore (1986) | 1986 | 847 | 26.47 | Boundary-Layer Meteorology |
| 12 | Zhao et al. (2005) | 2005 | 782 | 60.15 | Remote Sensing of Environment |
| 13 | Law et al. (2002) | 2002 | 780 | 48.75 | Agricultural and Forest Meteorology |
| 14 | Mu et al. (2007) | 2007 | 749 | 68.09 | Remote Sensing of Environment |
| 15 | Janssens et al. (2001) | 2001 | 687 | 40.41 | Global Change Biology |
| 16 | Wofsy et al. (1993) | 1993 | 685 | 27.4 | Science |
| 17 | Papale et al. (2006) | 2006 | 659 | 54.92 | Biogeosciences |
| 18 | Baldocchi (2008) | 2008 | 656 | 65.6 | Australian Journal of Botany |
| 19 | Goulden et al. (1998) | 1998 | 592 | 29.6 | Science |
| 20 | McMillen (1988) | 1988 | 569 | 18.97 | Boundary-Layer Meteorology |

Note: TC is the number of total citations of this article; TT is the temporal trends; CPY means citation per year.

## 3.7 Fund support

Source of funding could reflect R&D investment of a country or institute and how much attention do they pay to the specific field. In our investigation, 3073 publications out of 8297 could be tracked with their fund sponsors. Fig. 5 lists the top 15 sponsors of funding from the recognized 694 sponsors that support research based on eddy covariance. Among them, most publications have claimed that their research was supported by National Science Foundation (NSF) from the US, while National Natural Science Foundation of China came second. These are two largest sources of funds supported by government in the United States and China respectively. Other than these, US has two more sources that contribute to research based on eddy covariance, namely U.S. Department of Energy and NASA, while China has two extra funds from Chinese Academy of Sciences and National Basic Research Program of China. On the other hand, Natural Environment Research Council from UK turns out to be the fourth largest source of funding, and sponsors of funding from Canada, Finland, Germany, Japan and Turkey could all be tracked in the top 15 list, which reveals the international impact of research supported by eddy covariance.

**Fig. 5. Top 15 funding sponsors, 1981-2018. The number outside the brackets means how many papers were tracked to be supported by this funding. The percentage in the brackets is the proportion of research under this funding to the total research number with tracked funding.**

# 4. Research trends of eddy covariance

## 4.1 Knowledge structure and temporal trends

For high-frequency keywords occurring more than 30 times, we have drawn maps (Fig. 6) based on keyword co-occurrence relationships using VOSviewer software. Search keywords including "eddy covariance" were removed to make the network more informative, and keywords with the same meaning have been merged (e.g. "lue", "light-use efficiency" and "light use efficiency" were all merged into "light use efficiency"). Both cluster analysis and trend analysis are carried out to explore the historical trends of eddy covariance.

The author keywords represented by nodes are clustered broadly into five groups of topics (Fig. 6A): (1) NEE (red cluster), (2) ET (yellow cluster), (3) GPP (blue cluster), (4) Heat flux and energy balance (green cluster) and (5) Greenhouse gas (purple cluster). Considering temporal trends (Fig. 6B), we could find that in early days scholars were interested in carbon flux and heat flux, later carbon dioxide has gained more popularity as a global warming induced gas. In the next stage, keywords in the cluster of NEE (including NEE, carbon balance, respiration, photosynthesis, etc.) were topics that researches using eddy covariance first cared about, then keywords in the ET cluster considering water and heat flux have gained more attention. Later, the keywords from GPP cluster supported by various remote sensing products and machine learning models became more popular. At the same time, more and more new work considering topics of methane, greenhouse gas and climate change were emerging more often and gaining attention from researchers using eddy covariance.

(A)

(B)

**Fig. 6. Keyword co-occurrence network of eddy covariance. Only keywords with frequency of more than 30 are showed. The size of nodes reflects the occurrence number of the keywords. (A) Cluster analysis. Different color of nodes represents**

**different groups. (B) Temporal trends. The brighter the color of nodes, the more likely they were used as author keywords in recent years.**

## 4.2 Research hotspots

As carrier of key contents of a paper, keywords could well reflect why researchers have initiated their study and how they do it. Therefore, keyword analysis could explore the hot topics, including popular concepts and common methods, in the field. Moreover, by considering keywords in a temporal dynamical network based on their co-occurrence relations, we are able to find the most popular keywords in the investigated period as well as the potential hotspots in the future. This method has been stated and validated in our previous work(Huang and Zhao 2019), and is applied to calculate the popularity index and fitness index in this study.

Table 6 has listed top 20 popular keywords in publications related to eddy covariance. Most of them could be found in the maps created in Fig. 6. The most popular keyword is "evapotranspiration". Holding a PI of 1368.73, it has been included in 776 papers and co-occurred with 1634 unique keywords. Generally, keywords with high frequency might be more popular, but this is not an absolute requirement. For instance, "climate" has occurred only 143 times in publications, but it is more popular than keywords including "ecosystem respiration", "net ecosystem production", "carbon dioxide" and "energy balance" with larger or equal frequency, because according to its temporal trend it would be more likely for this topic to be mentioned more often in the future.

Concerning the potential hotspots, we could find some relatively rare keywords in the list (Table 7). These keywords could be largely classified into five groups: (1) New application in agriculture ("bioenergy", "switchgrass", "maize", "cropland"), (2) synthetic research with spatial analysis ("geographic location/entity", "landsat", "land-use change"), (3) methodology ("chlorophyll fluorescence", "lysimeter", "sensor", "land surface temperature"), (4) study sites ("china", "heihe river basin") and (5) conceptual terms ("atmosphere-land interaction", "carbon sink", "flux"). These keywords have a rather high fitness index though their keyword frequency might be low. They have the potential to gain more attention by eddy covariance researchers in the future. This results not only predict what might be hot topics in eddy covariance study in the future, but might also reveal the consensus of terms used by researchers over time.

**Table 6 Top 20 most popular keywords, 1981-2018 (ordered by PI)**

| Rank | Keyword | PI | TP | D |
|---|---|---|---|---|
| 1 | evapotranspiration | 1368.73 | 776 | 1634 |
| 2 | gross primary production | 977.86 | 430 | 972 |
| 3 | remote sensing | 667.23 | 279 | 736 |
| 4 | net ecosystem exchange | 579.99 | 342 | 781 |
| 5 | modis | 486.58 | 186 | 485 |
| 6 | light use efficiency | 483.31 | 168 | 447 |
| 7 | climate change | 447.75 | 143 | 454 |
| 8 | ecosystem respiration | 375.15 | 189 | 468 |
| 9 | net ecosystem production | 325.67 | 143 | 405 |
| 10 | carbon dioxide | 317.57 | 190 | 515 |
| 11 | energy balance | 304.04 | 199 | 479 |
| 12 | drought | 302.44 | 104 | 359 |
| 13 | water use efficiency | 275.33 | 106 | 293 |
| 14 | carbon cycle | 272.79 | 112 | 341 |
| 15 | photosynthesis | 271.28 | 159 | 463 |
| 16 | phenology | 255.55 | 77 | 269 |
| 17 | carbon flux | 243.59 | 98 | 308 |
| 18 | fluxnet | 241.59 | 75 | 300 |
| 19 | grassland | 241.23 | 91 | 285 |
| 20 | eddies | 238.44 | 56 | 169 |

Note: PI is the popularity index of the keyword; TP is the number of total publications containing this keyword; D, short for term "degree" in network science, represents how many distinct keywords have co-occur with this keyword.

**Table 7 Potential hotspots of eddy covariance (ordered by FI)**

| Rank | Keyword | FI | TP | D |
|---|---|---|---|---|
| 1 | atmosphere-land interaction | 3.84 | 18 | 47 |
| 2 | bioenergy | 3.8 | 16 | 68 |
| 3 | switchgrass | 3.73 | 15 | 48 |
| 4 | potential vorticity | 3.57 | 16 | 83 |
| 5 | maize | 3.56 | 34 | 107 |
| 6 | china | 3.51 | 16 | 44 |
| 7 | lysimeter | 3.44 | 17 | 57 |
| 8 | fluxes | 3.34 | 41 | 129 |
| 9 | land-use change | 3.34 | 13 | 68 |
| 10 | geographic location/entity | 3.31 | 7 | 38 |
| 11 | actual evapotranspiration | 3.31 | 31 | 100 |
| 12 | landsat | 3.29 | 26 | 97 |
| 13 | chlorophyll fluorescence | 3.26 | 7 | 29 |
| 14 | sensor | 3.24 | 6 | 37 |
| 15 | cropland | 3.22 | 24 | 73 |
| 16 | land surface temperature | 3.22 | 20 | 73 |
| 17 | heihe river basin | 3.21 | 14 | 41 |
| 18 | carbon sink | 3.2 | 25 | 94 |

Note: This table has included top 50 keywords according to FI excluding top 50 keywords ordered by PI. FI is the fitness index of the keyword; TP is the number of total publications containing this keyword; D represents how many distinct keywords have co-occurred with this keyword.

**4.3 Citation analysis**

For the 8297 articles based on eddy covariance, we have found 73,955 citations (excluding self-citation) citing these articles from 1981 to 2018. According to Scopus source list (https://www.scopus.com/sources, retrieved at March, 2019), these publications come from 261 different sub-subjects in 27 main subjects. The top 3 main subjects are Environmental Science (52.2%), Earth and Planetary Sciences (51.7%) and Agricultural and Biological Sciences (40.3%). That is to say, more than half of these publications that cited research based on eddy covariance could be categorized into Environmental Science or Earth and Planetary Sciences. Fig. 7A has listed the top 20 sub-subjects of the citing articles. Most of them belong to Atmospheric Science (18.0%), while 9.5% of publications are categorized as Ecology under Environmental Science. Still, 8.2% of these publications were recognized as research of Ecology, Evolution, Behavior and Systematics under Agricultural and Biological Sciences. Other important sub-subjects with a proportion larger than (or equal to) 6% include Water Science and Technology (8.7%), Soil Science (6.7%), Forestry (6.5%), General Earth and Planetary Sciences (6.2%), Global and Planetary Change (6.2%) and Environmental Chemistry (6.0%). From this picture we could find that eddy covariance technique, originated from a branch of atmospheric science known as microclimatology, has supported various subfields in basic science like ecology and environmental science, as well as applied science including agriculture and forestry. The temporal trend has told us that the main service target of eddy covariance before 2000 were scholars from Earth and Planetary Sciences (Fig. 7B), but Environmental Science has benefited more from EC studies after 2000. Overlapping the Environmental Science field, Agricultural and Biological Sciences has also paid more attention to eddy covariance studies over time.

(A)

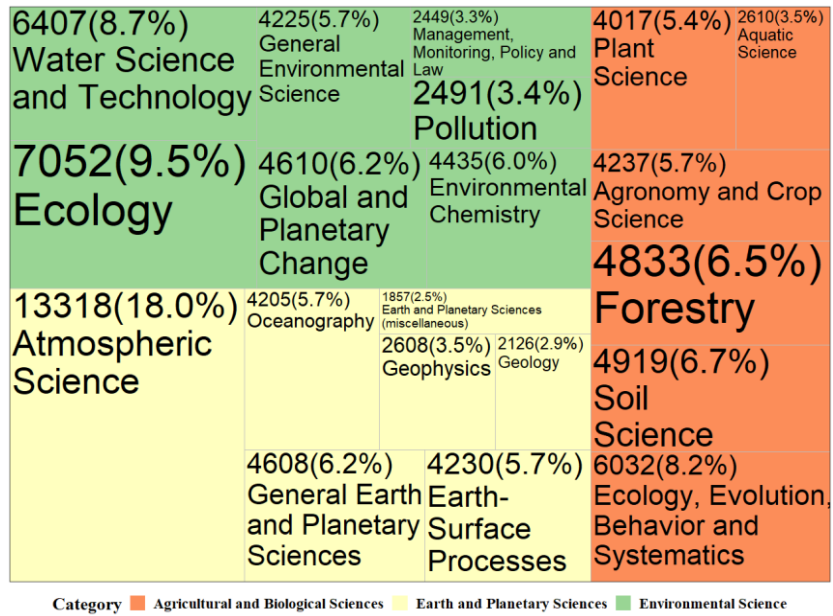

(B)

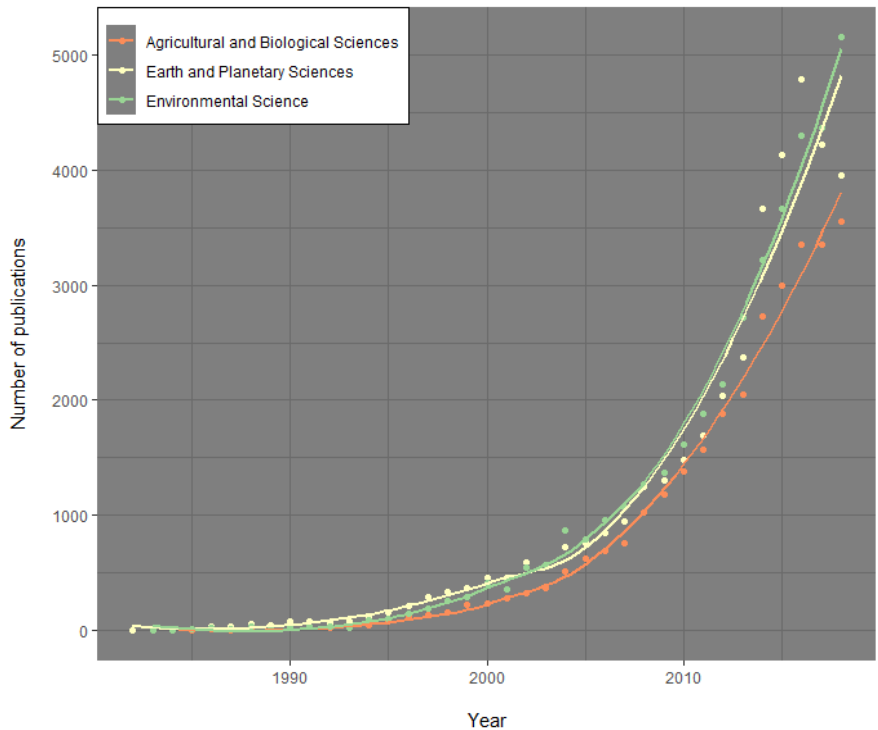

**Fig. 7. Citation analysis of eddy covariance. (A) Top 20 subjects of citations citing research based on eddy covariance, 1981-2018. The number outside the brackets is number of total publications classified as the sub-subject, while the percentage number inside is the proportion of publications classified as the sub-subject to**

the total publication. **(B) Temporal trends of citations. Each point represents how many citations could be classified a specific field, the solid fitted line is based on local polynomial regression fitting. Only top 3 fields are displayed in the plot.**

## 5. Collaboration among eddy covariance researchers

### 5.1 Development of collaboration in eddy covariance studies

To answer fundamental questions in the era of big science, collaboration is indispensable. In Fig. 8, we have quantified the extent of collaboration among eddy covariance researchers by measuring the author number, institution number and country number per paper. In general, we could find a significant rising trend of collaboration in the research community. At 1981, the author number of an article related to eddy covariance is 1.82 from 1.12 institutions by average. During 1990s, the author number per paper went up rapidly, and by the year of 2010 this number has risen to 6.08. At the same time, authors from different institutions are more willing to cooperate with each other. 2002 was the first year that the average institution number per paper broke through three (3.17), and this trend remained steady and reached nearly four (3.87) at 2018. Though not so obvious, the number of countries per paper is also on the rise. In 1980s and 1990s, there were fewer studies that collaborated by researchers from two countries. But in 21st century, collaboration across countries becomes more common and at 2018 the average country number per paper was 1.97. Interestingly, the standard deviation of average country number per paper is also increasing during the investigated period, indicating that while domestic research still dominates in general, a considerable proportion of studies supported by eddy covariance might demand more countries to get involved.

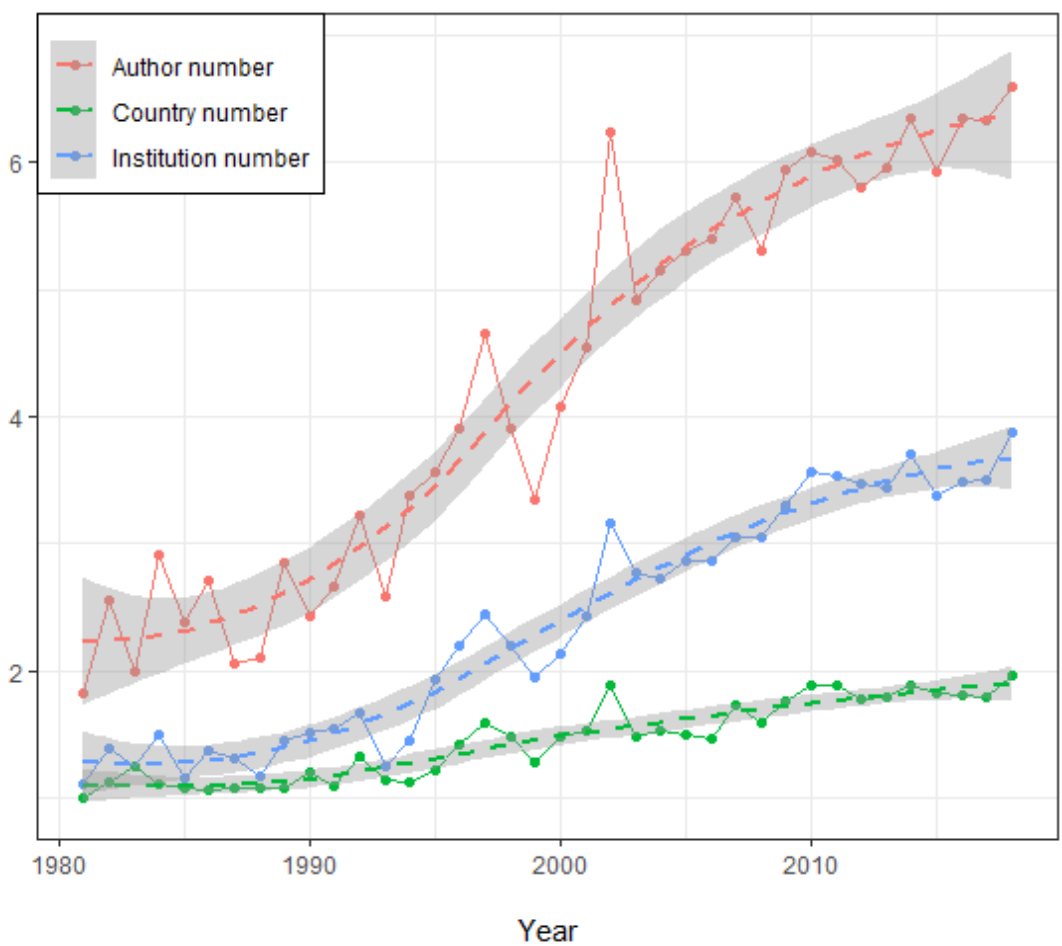

**Fig. 8. Development of collaboration in eddy covariance studies (1981-2018). The y-axis shows the annual number per paper (e.g. the red line represents the averaged author number per paper by year). Fitting dashed line and its fluctuation range is based on local polynomial regression fitting.**

## 5.2 FLUXNET: Unite researchers to a more solidary community

As more flux towers are established at different continents, these flux sites are linked together to form a large global network called FLUXNET. With its wide coverage, comparative and synthetic studies could be carried out to answer scientific questions at larger scale. As eddy flux sites are maintained by researchers from various institutes in different countries, collaboration is indispensable. In Fig. 9 we try to explore the effect of FLUXNET on collaboration network at international, institutional and individual level. Because FLUXNET was not established until ~1997, and the earliest literature tracked as FLUXNET eddy covariance study was published in 1998, here we include only papers published from 1998 to 2018 for this specific analysis. At international level, the collaboration network including FLUXNET research has 1 more node and 61 edges that the original network (Fig. 9A, Fig. 9B). This indicates that FLUXNET has attracted 1 more country to get involved in the whole investigated period (1981-2018), and 61 new connections between countries were made to link all the countries to do research together. Likewise, FLUXNET has brought 107 more institutions and 4205 more links in the institutional collaboration network. At individual level, FLUXNET has attracted 515 more researchers to join, and 14,629 co-authorship relations are formed when the FLUXNET research is introduced.

With more nodes and edges brought by FLUXNET, the structures of collaboration network change accordingly. Fig. 9 shows that the collaboration network would become more "compact" with FLUXNET research included, small clusters would merged into larger clusters (e.g. large clusters in institutional collaboration network tend to include more nodes when FLUXNET research are considered). The average degree (AD) of nodes increases in networks at all three levels. On the whole, FLUXNET has made collaboration networks more solidary, with more countries, institutions and researchers connecting closely to each other.

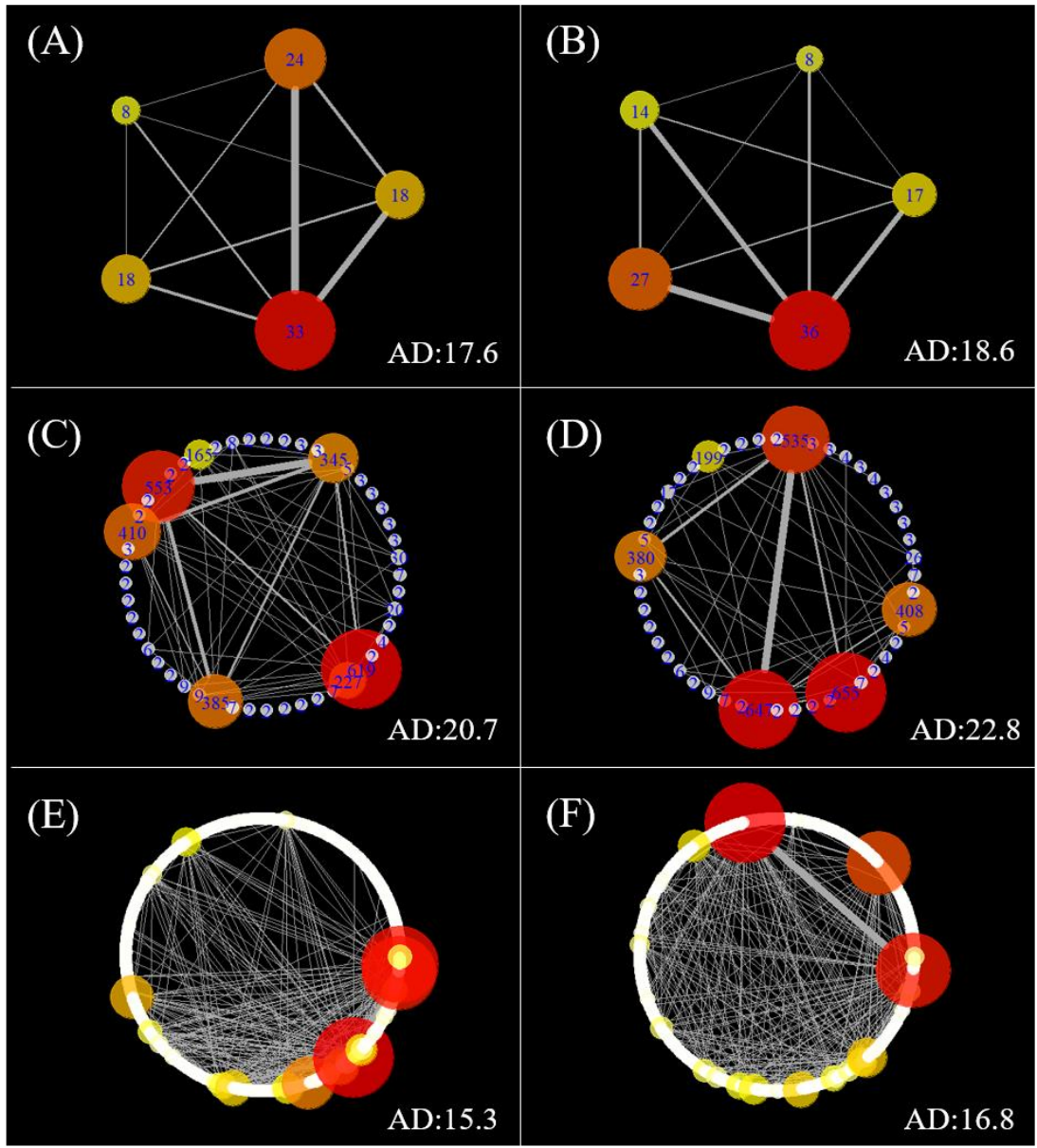

**Fig. 9. Comparison of collaboration network without or with research supported by FLUXNET from 1998 to 2018 at three levels (international, institutional and individual). (A) International collaboration network without FLUXNET research. (B) International collaboration network with FLUXNET research. (C) Institutional collaboration network without FLUXNET research. (D) Institutional collaboration network with FLUXNET research. (E) Author collaboration network without FLUXNET research. (F) Author collaboration network with FLUXNET research. AD is the average degree of nodes. Blue labels in sub-figures A, B, C, D show number of nodes in cluster, they are not displayed in author collaboration network because there are too many clusters.**

## 5.3 Impact analysis of FLUXNET research

With the support of FLUXNET, scientists are now able to utilize eddy covariance measurements all around the world to upscale their research and answer questions that are more general. These researches might gain wider attention and benefit the scientific community at broader scale. Here we want to evaluate the impact of FLUXNET research by comparing it with other research supported by eddy covariance. As the fundamental metric for calculation of H-index, Impact Factor, CiteScore, etc., citation count is a vital indicator to measure the influence of scientific research. Therefore, in our study we use this metric to estimate and compare the impact of FLUXNET research and others. This analysis is based on publications from 1998 to 2018 for the same reason in the former section. In Fig. 10, we could find that overall FLUXNET research have greater impact than other research supported by eddy covariance. This might not be so obvious on the publication year. But on the first year after publication, a research supported by FLUXNET could get 5.28 citations by average, while other research gets 3.28 citations. In addition, this impact gap would enlarge overtime. At the third year after publication, work supported by FLUXNET gets 20.9 accumulative citations per paper, while other research gets 12.1. After five years' time, the average citation count of FLUXNET research is 39.2, that's ~1.8 times higher than other research (21.8).

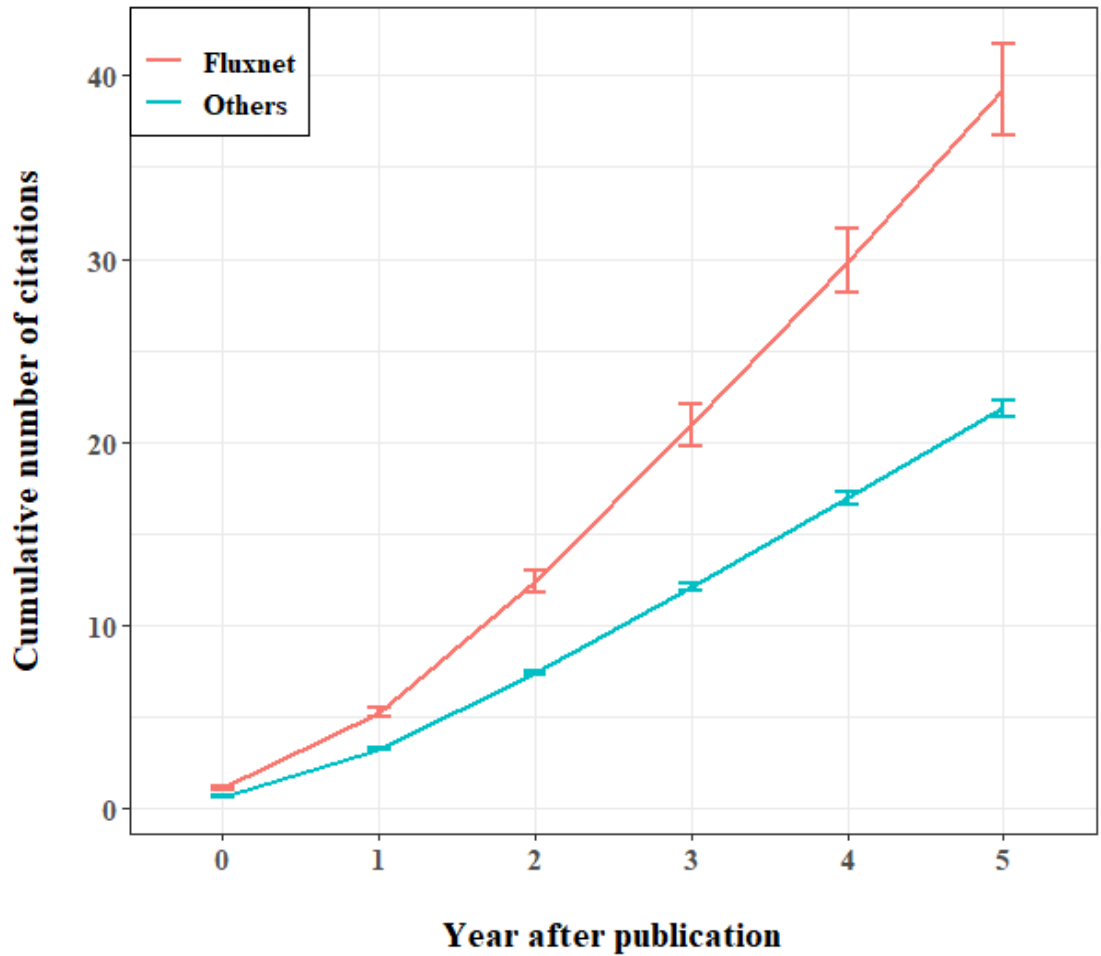

**Fig. 10. Impact analysis of FLUXNET research. The error lines show the standard errors of the groups.**

## 6. Discussion

### 6.1 Application of eddy covariance in ecology

Today, eddy covariance technique has become a common useful method to measure

$CO_2$ flux between terrestrial ecosystem and atmosphere. It is, however, neither the first

nor the only method to do such measurement. First experiments that try to acquire $CO_2$ exchange were based on flux-gradient method(Inoue et al. 1958, Lemon 1960, Monteith and Szeicz 1960), which proved to be problematic when applied in forest ecosystems(Raupach 1979, Simpson et al. 1998). Other optional methods include enclosure, cuvette and chamber, etc. While having their own merits in different circumstances, they suffer from various drawbacks such as introducing extra bias, being labor intensive and hard to upscale(Baldocchi 2003). On the other hand, developed by physicists based on fluid dynamics(Reynolds 1895), eddy covariance was first applied on studies of momentum transfer(Swinbank 1951). Before $21^{st}$ century, most attention gained by eddy covariance comes from atmospheric sciences (Table 4), and studies were usually driven by questions in Earth and Planetary Sciences (Fig. 7B). During this period, many technical questions were raised and discussed, and scientists were conducting various experiments to explore the possibility for application. $CO_2$ exchange measurement is one of those applications that drive the development of eddy covariance technique. For instance, devices like sonic anemometers, open-path $CO_2$ sensors, infrared spectrometers had been created to measure $CO_2$ more accurately in longer periods(Jones et al. 1978, Wofsy et al. 1993). Since then, eddy covariance, as a technique measuring $CO_2$ exchange and beyond, had moved to a relatively mature stage.

By now, eddy covariance has gained wide popularity from researchers of extensive fields, such as ecology, forestry, oceanography and environmental science (Fig. 4). Compare eddy covariance system to a tree, micrometeorology would be its stems with roots at fluid dynamics, and the precision measurement methods were leaves refreshing over time. Just like the flowers are blooming and fruits are produced as the tree grows, numerous applied studies are flourishing with the advent of mature technical system. In Fig. 6, we have drawn a picture to show these flowers and fruits. It is not difficult to find that major applications of eddy covariance are surrounding the field of ecology, "net ecosystem exchange" and its synonyms were popular keywords in eddy covariance papers at the very beginning. The reason is that global carbon cycle, as an important component of biogeochemical cycle, had always been concerned by ecologists. And this process becomes a vital focus for researchers when greenhouse gas induced climate change receives more and more attention from the general public. Eddy covariance, providing a direct measure of net $CO_2$ exchange between ecosystem and atmosphere, serves as a powerful tool to track the footprint of carbon across the canopy-atmosphere interface. Different from other traditional means, it provides a scale-appropriate method at longer periods and broader space(Baldocchi 2003).

Apart from carbon, water and energy land-atmosphere exchanges are also important drivers in global climate system, which could be quantified and estimated as water flux and heat flux in eddy covariance measuring system. From the view of ecosystem and beyond, processes of carbon cycle, water cycle and energy flow are coupling with each other in nature, so it is difficult and inappropriate to study them separately. Later, "evapotranspiration" has become another research hotspot among ecologists using eddy covariance. Consisted of evaporation from land surface and transpiration from plant leaves, evapotranspiration (ET) is an important ecological process that links water, carbon and energy cycles. It serves as a key concept for other research topics, such as water budget, water use efficiency, energy balance, energy budget, etc. Recent decades have seen striking progress of ET science and applications, including ecosystem functioning, agriculture production and water management, but to upscale these achievements from local to global remains to be developed(Fisher et al. 2017).

## 6.2 Remote sensing: A driving factor to upscale eddy covariance

The ultimate goal of eddy covariance research community is to produce "everywhere, all of the time" information of flux(Baldocchi 2014), but it is unpractical to build and maintain so many flux towers across the globe at current state. Remote sensing is the very technique to fill this gap. As a invaluable tool to capture information on Earth without direct physical contact, it has been widely used in ecosystem evaluation and biodiversity conversation, such as monitoring, mapping, modelling and decision support(Horning et al. 2010, Pettorelli et al. 2014, Pettorelli et al. 2018, Vaz et al. 2018, Zellweger et al. 2019). Combined with remote sensing products and latest machine learning methods, flux maps could be produced at a much larger area. In this way, flux data could be scaled up from sites to region(Ouimette et al. 2015, Wang et al. 2016), from regions to continent(Xiao et al. 2011), and from continents to global(Beer et al. 2010). Supported by empirically upscaled fluxes, these maps could be used to monitor temporal changes in terrestrial fluxes, provide independent data to evaluate terrestrial biosphere models, or help test and improve current models(Ichii et al. 2018, Running et al. 1999). The prospect of integration of eddy covariance and remote sensing is promising, but challenge remains because related programs would demand collaboration across the globe and optimization of resources allocation. If eddy flux data are too sparse in space, they might be less helpful in practical application (Chevallier et al. 2012).

### 6.3 Data sharing: History, current state and future prospects

FLUXNET is a global network of eddy covariance flux measurements. The project has served to unite eddy covariance researchers (Fig. 9) and provide a platform integrating the relatively isolated flux data from all over the world(Baldocchi et al. 2001). In historical development, FLUXNET had received supports from three main sources from the US, which were NASA, NSF and U.S. Department of Energy chronologically (https://fluxnet.fluxdata.org/about/history/). These funds could all be tracked in our results (Fig. 5). Started at around 1997 with support from NASA, FLUXNET was first initiated to provide ground support for satellite products produced by Terra and Aqua, namely Earth Observing System (EOS) and Moderate Resolution Imaging Spectroradiometer (MODIS). By then, regional flux networks like AmeriFlux, AsiaFlux and EuroFlux had been established. The role of FLUXNET was to unite the regional networks and build an integrated database for access, so as to facilitate comparison and synthesis studies. Later, receiving support from NSF since 2007, the ever-growing database contributed by the extensive network had attracted lots of computer scientists, both from business and academia, to develop a new data system for the massive complex flux data. During this period, the La Thuile Database was produced and released. This dataset contains flux data of carbon, water and energy as well as related meteorological information from 252 sites. Together with the new interactive data system, the database has greatly promoted synthesized studies supported by eddy covariance, such as carbon exchange, water exchange, model validation and model parameterization. In 2014, FLUXNET received funds from Department of Energy, USA. At current stage, FLUXNET would make more efforts to provide flux data at regular basis, and value-added data products would be made to meet the new challenges of global biogeoscience community.

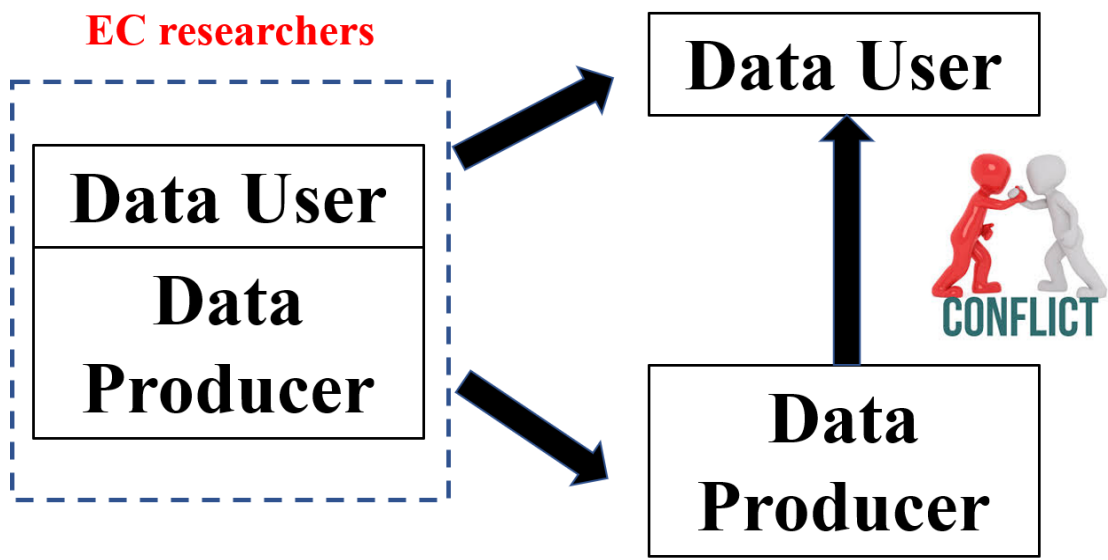

**Fig. 11. Separation of data user and data producer among EC researchers.**

Though the overall data-intensive eddy covariance research is heading for a more collaborative science (Fig. 8), the actual situation of data sharing is not that encouraging. It was estimated that only 8% of eddy covariance researchers uploaded their published data on public data portals(Dai et al. 2018). For these 8% researchers, only 64% of them provided directly downloadable datasets online, while others only offer their metadata instead of raw data. Reasons for this phenomenon might include being logistically inconvenient, desiring to use data for future publication, lack of sufficient credit, etc., which has been discussed by previous studies(Bond Lamberty 2018, Parr and Cummings 2005). To conclude, although data sharing is beneficial and efficient in the general interests of scientific community, in the current culture and evaluation system it is uneconomical for individual scientists and institutions to make their data openly available. In cognition of most researchers, data users earn much more credits than data sharers (producers), who might also become the potential data users in the future. Previously, eddy covariance research was often carried out by one small group. Therefore, data producers and data users were usually the same researchers without conflict of interest (Fig. 11). Research achievements are evaluated as a whole, neglecting the subdivisions of study process such as data production, processing, modelling and visualization. But in the new era of big data, research paradigm for collaborative science has separated data producers from data users. It is a great advancement, yet not accomplished until the advent of new evaluation system driven by policy. For instance, our results reveal that research supported by FLUXNET are more likely to gain greater impact in the following five years (Fig. 10), but we fall short of methods to quantify the contributions provided by these data sharers. Likewise, it was argued that positive correlations might exist between data sharing and academic impact(Dai et al. 2018). This academic impact, however, was determined by citation and cooperation, but not how much high-quality data they have shared or how many times their data products have been reused. Efforts have been made to adequately reward the contributions of data sharers, such as enlisting as coauthors, publication of "data papers" and setting up incentive systems. They work locally, in the short term, but could not solve the problem fundamentally. To meet this challenge, a more comprehensive quantitative evaluation system for data sharing should be established and promoted in the general academia for the long run. The measurement unit in academics is usually paper, and a paper has author list to track researchers, affiliation list to track institutions, reference list to track citations and fund list to track financial support. It is just a matter of time, we believe, for every paper to have a data list tracking the contribution of data sharers. By then, scientists will be able to cooperate freely with each other via data sharing, just like how they do with literature citation today.

## Acknowledgement

All our summary statistics and R codes are freely available on request. Our access to the Scopus and Web of Science comes through a contract with that forbids redistribution of their database. Researchers who desire to get the raw data and replicate our analytics could obtain paid subscription to Thomson Reuters and Elsevier. Here, we thank Ludo Waltman for his professional advice on network visualization using VOSviewer. Also, we thank Dennis Baldocchi for his helpful comments on some of our results. This study was supported by grant from the National Key Research and Development Program of China (Grant No. 2018YFD0900806) and the National Natural Science Foundation of China (Grant No. 31170450).